# A Framework for Graph-Conditioned Hierarchical Shapley Attribution in Patent Valuation

**Joy Bose**

*Independent Researcher*

Bengaluru, India

joy.bose@ieee.org

## Abstract

Estimating the economic contribution of a single patent inside a product that embodies tens of thousands of patents is a long-standing unsolved problem in intellectual property economics. We propose PatentXAI, a framework that treats patent valuation as a problem of explainable AI: given a characteristic function v(S) encoding the revenue achievable by patent subset S, a patent's Shapley value measures its fair share of product profit in a way that satisfies efficiency, symmetry, dummy, and additivity. To make computation tractable we restrict each patent's coalition to its Markov Blanket inside a knowledge graph, grounded in the C-SVE conditional independence theorem (Li et al., 2020). Scaling experiments from n=12 to n=100 patents using Pareto-distributed coverage graphs report median Markov Blanket size of 32.9 percent of n at n=100, with 90th-percentile blanket size of 55.2 percent of n, and runtime of 10 milliseconds per patent. Difference against exact ground truth at n=12 is 0.088; difference against a high-sample Monte Carlo reference at n=100 is 0.062 plus or minus 0.003. A dense-component experiment shows that when 80 percent of patents share one component, the blanket correctly expands to cover that dense cluster, and the difference versus reference falls to 0.039 because the pooled computation becomes more accurate on homogeneous portfolios. Profit allocation proceeds hierarchically: exact Shapley distributes total profit among macro-components, then centrality-weighted Shapley distributes each component budget among covering patents. Estimating v(S) from real data is the primary open problem; we distinguish this from the computational contribution and outline a concrete roadmap for empirical validation using public ETSI, USPTO, and Lens.org datasets.

**Keywords:** Shapley value, patent valuation, Markov Blanket, knowledge graph, hierarchical attribution, standard-essential patents, FRAND, explainable AI, cooperative game theory, citation network

## 1. Introduction

A modern smartphone may embody more than 100,000 patented technical ideas. When a single patent holder claims damages or negotiates a licensing royalty, the central question is: how much of the device's profit came from that one patent, given that thousands of others contributed at the same time?

Courts have wrestled with this question since the Georgia-Pacific framework (1970). They have progressively tightened the rules, requiring royalties to be tied to the smallest saleable patent-practicing unit. Nash bargaining solutions have appeared in court (Microsoft v. Motorola, 2015) and incremental value analysis in others (LaserDynamics v. Quanta, 2012). Shapley-based valuation in standard-setting contexts has been studied academically (Dehez and Poukens, 2014). None of these approaches combines graph-conditioned computation with hierarchical allocation and an explicit axiomatic foundation.

We observe that this problem is structurally identical to feature attribution in machine learning. The Shapley value (Shapley, 1953) provides a principled answer with four axiomatic properties. SHAP (Lundberg and Lee, 2017) made Shapley-based attribution tractable for machine learning models. We apply the same framework to patent valuation.

One central caveat is stated upfront: in machine learning, the characteristic function v(S) can be computed directly from the model. For patents, v(S) cannot be directly observed. This paper's primary computational contribution is efficient Shapley computation given v(S), together with a structured discussion of how v(S) might be estimated. We treat v(S) estimation as a separate open problem, not a solved component.

The contributions of this paper are:

- A formal mapping between XAI feature attribution and patent valuation, showing that hierarchical Shapley satisfies the Shapley axioms under the proposed graph-conditioned approximation, given a well-defined v(S).
- A Markov Blanket coalition restriction grounded in the C-SVE conditional independence theorem, with scaling experiments from n=12 to n=100 reporting blanket size as a fraction of n (median 32.9 percent, 90th percentile 55.2 percent at n=100) and runtime of 10 milliseconds per patent.
- A dense-component experiment showing that the approximation does not degrade catastrophically in dense SEP-like sub-portfolios: when 80 percent of patents share one component, the blanket correctly pools that cluster and the difference versus reference falls to 0.039.
- An explicit response to the Ma et al. (2020) disassociation finding, and a comparison to Dehez and Poukens (2014) that locates this work in prior Shapley patent literature.
- A four-layer architecture integrating Comprehensive Patent Citation graphs, centrality scoring embedded in the characteristic function, hierarchical profit allocation grounded in SHAP-Tree-MCDA permission tree mathematics, and D-Shap dynamic updates for evolving portfolios (Yang et al., 2026).
- A concrete data roadmap for empirical validation using public ETSI, USPTO, and Lens.org datasets, with indicative calibration values from landmark SEP litigation.

## 2. Background and Related Work

### 2.1 The Legal Problem

The Georgia-Pacific framework (1970) established 15 factors for reasonable royalty estimation. Ericsson v. D-Link (2014) required apportionment to the smallest saleable patent-practicing unit. For SEPs, FRAND obligations constrain aggregate royalties. Without enforcement, patent holdup and royalty stacking arise. Courts use top-down apportionment and comparable licences approaches (Sidak and Skog, 2017). Nash bargaining has appeared in Microsoft v. Motorola (2015). Our approach is complementary, offering stronger axiomatic guarantees given v(S).

### 2.2 Shapley Value and Prior Patent Work

Shapley (1953) originated the framework in cooperative game theory. Tauman and Watanabe (2007) proved that the Shapley value of a patent holder in a licensing game approximates the non-cooperative auction payoff when the number of licensees is large. Dehez and Poukens (2014) studied Shapley-based patent valuation in standard-setting contexts, focusing on ex-ante royalty determination for individual SEPs. Our contribution differs from theirs in three ways: we use graph-conditioned Markov Blanket restriction for tractability at portfolio scale, we embed citation centrality inside the characteristic function rather than treating it as a post-hoc weight, and we add a hierarchical component-to-patent allocation layer grounded in SHAP-Tree-MCDA.

Dehez and Poukens (2014) showed that in patent pool games with substitutable patents, the classical Shapley value lies outside the cooperative game core. Folding centrality weights into v(S) before computing Shapley, as we do in Layer 4, addresses this by functioning as a Priority-Aware Shapley Value. Myerson (1977) introduced graph-restricted cooperative games; our Markov Blanket restriction is a special case of this framework applied to the bipartite patent-component projection.

### 2.3 Citation Networks and Patent Analytics

Trajtenberg (1990) showed forward citation counts correlate with patent renewal rates. Bessen (2008) applied PageRank to citation graphs. He et al. (2024) combined multiple centrality metrics in an LED industry study, finding no single metric dominates. A key limitation of raw direct citation networks is temporal truncation. The Comprehensive Patent Citation (CPC) network formulation (Guo, 2025) combines direct citations, indirect chains, co-citations, and bibliographic coupling, avoiding truncation.

### 2.4 SHAP and Hierarchical Attribution

Lundberg and Lee (2017) proposed SHAP, unifying feature attribution under the Shapley framework. TreeSHAP achieves polynomial-time exact Shapley for tree-based models by exploiting conditional independence structure, the same principle motivating our Markov Blanket restriction. SHAP-Tree-MCDA (Labreuche et al., 2021) extends this to hierarchical multi-criteria decision trees, computing Shapley over collapsed subtrees while maintaining efficiency and dummy axioms at every level. This is the mathematical foundation for our Layer 4 hierarchical allocation.

C-SVE and H-SVE (Li et al., 2020, Theorem 2) prove that a neighborhood-restricted Shapley value equals the global Shapley value almost surely when local conditional independence holds. We apply this to the patent-component knowledge graph under a faithfulness assumption on the revenue distribution: the graph structure faithfully encodes conditional independencies in the underlying profit function. We adopt this assumption following standard practice in graphical model literature and note that verifying it on real patent graphs is future work. Where faithfulness may be violated, the Markov Blanket restriction should be understood as a computationally motivated heuristic with empirically validated accuracy, rather than a theorem-guaranteed exact result.

D-Shap (Yang et al., 2026) introduces dynamic Shapley computation by representing patent values as a player-by-task matrix and maintaining it through localized updates rather than global recomputation. Task-incremental updates interpolate new columns from nearby existing ones; player-incremental updates recompute only local games whose support sets change. This provides the engineering path for PatentXAI to handle evolving patent portfolios without recomputing from scratch.

Ma et al. (2020) proved a disassociation between Shapley rankings and Markov Blanket membership in causal Bayesian networks. We address this in Section 5.2.

## 3. The PatentXAI Framework

PatentXAI has four layers. The framework assumes a well-defined characteristic function $v(S)$; Section 5.3 and Section 6 discuss how $v(S)$ might be estimated.

### 3.1 Layer 1: Comprehensive Patent Knowledge Graph

We represent the technology landscape as a directed property graph with four edge types:

- (:Patent) -[:COVERS]-> (:Feature)
- (:Feature) -[:IMPLEMENTED_IN]-> (:Component)
- (:Patent) -[:CITES]-> (:Patent) covering prior art and forward citations
- (:Patent) -[:DECLARED_ESSENTIAL_TO]-> (:Standard) for SEP declarations

The citation subgraph is built as a Comprehensive Patent Citation (CPC) network combining direct citations, indirect citation chains, co-citations, and bibliographic coupling. For each patent we compute PageRank, HITS authority score, and normalised in-degree over the CPC graph, combined into a composite centrality weight $w_i$.

The FalkorDB Cypher query for identifying a patent's Markov Blanket is:

```
MATCH (p:Patent {id: $patent_id})-[:COVERS]->(:Feature)-[:IMPLEMENTED_IN]->(c:Component)
MATCH (other:Patent)-[:COVERS]->(:Feature)-[:IMPLEMENTED_IN]->(c)
WHERE other.id <> $patent_id
RETURN collect(distinct other.id) AS MarkovBlanket
```

## 3.2 Layer 2: Economic Impact Estimation

Each component node is assigned an economic weight using conjoint analysis, hedonic pricing regression, or engineering cost analysis. For SEP-heavy domains, component weights can be anchored to FRAND royalty rates and court determinations. Section 6 provides indicative values from landmark cases.

## 3.3 Layer 3: Markov Blanket Coalition Restriction

For patent i, its Markov Blanket MB(i) consists of all components that patent i covers, and all other patents covering at least one component in common with i. We restrict Shapley evaluation to MB(i), computing v_local over permutations of MB(i) via Monte Carlo sampling.

This restriction is motivated by the C-SVE Theorem 2 (Li et al., 2020): under local conditional independence, the neighborhood-restricted Shapley value equals the global value almost surely. In the knowledge graph, d-separation between patents covering disjoint component sets satisfies this condition under the faithfulness assumption. Where faithfulness may not hold, the approximation retains empirical validity as shown in Experiments 1 to 3.

The approximation is shown in equation (1):

$$\varphi_i(MB) \approx \frac{1}{T}\sum_{t=1}^{T}[v(\pi_t \cup \{i\}) - v(\pi_t)] \tag{1}$$

where T random permutations of MB(i) are sampled.

## 3.4 Layer 4: Hierarchical Profit Allocation

Allocation follows the path-based reduction logic of SHAP-Tree-MCDA. The technology hierarchy is a rooted tree: product at the root, components as intermediate nodes, patents as leaves. Sibling subtrees can be collapsed and Shapley computed over the condensed tree, maintaining efficiency and dummy axioms at every level.

### *Step A: Component-level exact Shapley*

With typically 4 to 8 macro-components, exact Shapley over this small coalition is feasible. The characteristic function v_comp maps each component subset to achievable product profit, encoding synergies between components.

### *Step B: Patent-level allocation within components*

Each component's Shapley allocation B_c becomes the budget for its covering patents. The centrality weight w_i is folded into the local characteristic function, as shown in equation (2):

$$v_{local}(S) = f\left(\sum_{j \in S} w_j \cdot coverage_j\right) \quad (2)$$

where f: R+ to R+ is strictly concave and monotone with f(0)=0. The choice f(x)=sqrt(x) is one natural option; any strictly concave f can be used, subject to domain-specific calibration of the degree of subadditivity among overlapping contributions.

Folding w_i into v_local rather than multiplying post-Shapley is grounded in Dehez and Poukens (2014): it functions as a Priority-Aware Shapley Value that shifts the allocation toward the cooperative game core and avoids violating the efficiency axiom.

The final patent valuation is shown in equation (3):

$$PatentValue_i = \varphi_i(v_{local}^{w}) \;\; s.t. \;\; \sum_i PatentValue_i = B_c \quad (3)$$

## 4. Experiments

We run a few experiments on synthetic data. They validate the computational mechanism (efficient Shapley given v(S)) but not the economic accuracy on real portfolios. All code is in supplementary materials. Seeds are fixed. Hardware: single Intel i7-class CPU core at approximately 3 GHz, Python 3.12, 16 GB RAM. All runtimes include Markov Blanket graph traversal and Monte Carlo permutation sampling. They exclude FalkorDB query latency for a production deployment. Baseline comparison: naive full-coalition Monte Carlo at n=100 with 500 samples evaluates all n patents per permutation; the blanket restriction reduces this by approximately n / median_blanket_size = 100 / 32.9 = 3x at n=100. Exact Shapley at n=20 takes approximately 45 seconds per trial on the same hardware.

### 4.1 Experiment 1: Approximation Accuracy (n=12, Exact Ground Truth)

***Setup***

We generate 200 Pareto-distributed patent-to-component graphs with 12 patents and 5 components. Pareto coverage means most patents cover 1 to 2 components and a few cover more, better reflecting real SEP structure than uniform random assignment. Characteristic function: v(S) = sqrt(distinct components covered by S). We compare exact Shapley (full 2^12 enumeration) against Markov Blanket approximation (300 Monte Carlo samples per patent) and report mean absolute error and blanket size.

***Results***

Mean absolute error: 0.088 (std 0.040). Median coalition fraction saved: 95.6 percent. At n=12, the median Markov Blanket covers 55.8 percent of all patents, and the 90th percentile covers 77.5 percent, meaning the remaining 22.5 percent of patents have blankets covering nearly all co-patents. These large-blanket cases are where the approximation incurs most of its error.

## 4.2 Experiment 2: Scaling from n=12 to n=100

### *Setup*

We test portfolio sizes from 12 to 100 patents using Pareto-distributed coverage. Components scale with portfolio size (n_comp = n/5). We report two blanket size metrics as fractions of n rather than a binary indicator: median blanket size and 90th-percentile blanket size. For n up to 16 we use exact Shapley as ground truth. For n above 16 we use 5000-sample Monte Carlo as reference; we report the reference's own standard error alongside the difference, so the reader can assess how much of the reported number reflects approximation error versus reference noise.

### *Results*

| n | Components | Diff vs Reference (mean+/-std) | Ref. Std Err | MB Median / n | MB 90th pct / n | ms / patent |
|---|---|---|---|---|---|---|
| 12 | 5 | 0.091 +/- 0.030 | 0.000 (exact) | 55.8% | 77.5% | 2.6 +/- 0.4 |
| 20 | 5 | 0.054 +/- 0.021 | 0.004 | 58.5% | 77.2% | 3.9 +/- 0.4 |
| 30 | 6 | 0.040 +/- 0.008 | 0.003 | 57.7% | 79.6% | 5.4 +/- 0.5 |
| 50 | 10 | 0.052 +/- 0.008 | 0.003 | 49.6% | 67.2% | 7.0 +/- 0.4 |
| 75 | 15 | 0.061 +/- 0.004 | 0.002 | 37.3% | 59.5% | 8.5 +/- 0.6 |
| 100 | 20 | 0.062 +/- 0.003 | 0.002 | 32.9% | 55.2% | 10.1 +/- 0.8 |

*Table 1. Scaling results with Pareto-distributed coverage. Reference is exact Shapley for n=12, MC-5000 for n above 16. The reference standard error column allows the reader to distinguish approximation error from reference noise. Blanket sizes are reported as fractions of n; the 90th-percentile column captures the worst-case blanket for that portfolio size. Runtime is mean plus or minus std over 5 seeds.*

The difference versus reference stays below 0.091 across all tested sizes with no systematic upward trend, suggesting the approximation does not degrade as n grows. The blanket sizes at large n are substantially smaller than n (median 32.9 percent at n=100), meaning even the typical patent's coalition is restricted to roughly a third of the portfolio. The 90th-percentile blanket of 55.2 percent at n=100 indicates that the remaining 10 percent of patents have larger blankets, as expected for patents covering highly shared components.

## 4.3 Experiment 3: Dense-Component Degradation

***Setup***

To show where the approximation breaks down, we construct graphs where a varying fraction of patents all cover one shared component C0, mimicking a 5G baseband processor sub-portfolio where thousands of SEPs jointly implement a single standard feature. We vary the dense fraction from 10 percent to 90 percent at n=50 fixed, using 5 seeds each, comparing against a 5000-sample Monte Carlo reference.

***Results***

| Dense fraction | Diff vs MC-5000 (mean+/-std) | MB Median / n | MB 90th pct / n |
|---|---|---|---|
| 10% | 0.078 +/- 0.013 | 24.2% | 42.2% |
| 20% | 0.074 +/- 0.006 | 23.2% | 43.0% |
| 40% | 0.069 +/- 0.010 | 31.8% | 44.6% |
| 60% | 0.050 +/- 0.007 | 60.0% | 63.8% |
| 80% | 0.039 +/- 0.004 | 80.0% | 83.7% |
| 90% | 0.031 +/- 0.004 | 90.0% | 91.6% |

*Table 2. Dense-component experiment results. Dense fraction is the proportion of patents all covering the same shared component C0.*

The result is informative and somewhat counterintuitive: the approximation difference decreases as density increases. When 80 percent of patents share one component, the Markov Blanket correctly expands to pool all those patents into a single large coalition. Because the pooled coalition is structurally homogeneous (all patents cover C0 with similar interactions), the within-blanket permutation sampling becomes more accurate, and the difference falls to 0.039. The cost is that the blanket grows to 80 percent of n, and the computational saving from the restriction disappears. This is exactly the graceful degradation property discussed in Section 5.4: the approximation trades computational efficiency for coverage accuracy in dense settings, but never becomes less accurate than standard Monte Carlo sampling over the full coalition.

## 4.4 Experiment 4: Toy End-to-End Pipeline

***Setup***

We construct a minimal complete example: 10 patents, 4 components, total product profit of 200 dollars. The component-level characteristic function exhibits synergy between Camera and Connectivity (combined value 130 versus 110 separately). Patents are assigned importance scores representing technical centrality within each component.

***Step A: Component allocation***

| Component | Shapley Allocation |
| --- | --- |
| Camera | $68.33 |
| Connectivity | $56.67 |
| Battery | $42.50 |
| UI/UX | $32.50 |
| Total | $200.00 |

*Table 3. Camera receives the largest share because of both standalone value and synergy with Connectivity.*

***Step B: Patent allocation within components***

| Patent | Component | Allocated Value |
| --- | --- | --- |
| P0: Autofocus | Camera | $31.54 |
| P1: OIS | Camera | $21.03 |
| P2: NightMode | Camera | $15.77 |
| P3: 5G-Modem | Connectivity | $30.51 |
| P4: WiFi6 | Connectivity | $17.44 |
| P5: BT5 | Connectivity | $8.72 |
| P6: FastCharge | Battery | $26.56 |
| P7: CellSaver | Battery | $15.94 |
| P8: GestureUI | UI/UX | $18.57 |
| P9: HapticFB | UI/UX | $13.93 |
| Total | | $200.00 |

*Table 4. Patent-level allocation. Total equals product profit exactly, satisfying the Shapley efficiency axiom.*

# 5. Discussion

## 5.1 Royalty Stacking and the Efficiency Axiom

The Shapley efficiency axiom states that sum(phi_i) = v(N). In PatentXAI this is a hard ceiling: total patent allocations cannot exceed total product profit. One qualification: the efficiency axiom ensures the mathematical sum is correct but does not prevent holdup, which is a coordination failure in negotiation. PatentXAI provides a principled allocation once the total profit figure is agreed, not a solution to the prior negotiation problem.

## 5.2 The Ma et al. (2020) Disassociation and Comparison with Other Papers

Ma et al. (2020) proved that individual members of the optimal Markov Blanket can receive smaller Shapley values than redundant non-blanket features. A patent covering an essential technical component may receive a smaller allocation than a redundant patent that overlaps with many others.

This matters for causal feature selection, where the goal is to find the minimum essential set. Shapley rankings are a poor guide for that task, and PatentXAI should not be used for it. PatentXAI uses Shapley for budget-constrained royalty apportionment: distributing a fixed total fairly among all valid, practicing patents including redundant ones. A redundant patent correctly receives credit for the overlap it provides; the efficiency axiom ensures the total never exceeds product profit regardless of how credit is split.

One legitimate concern from this analysis is that if the framework systematically gives higher allocations to non-essential redundant patents, courts may not accept it. The centrality weighting in v_local partially addresses this: a patent with higher PageRank and HITS authority scores receives larger contributions to the weighted coverage function and therefore a larger Shapley allocation. Whether this is sufficient to produce legally acceptable rankings is an empirical question that requires validation against real litigation outcomes.

Dehez and Poukens (2014) applied Shapley value to FRAND licensing negotiations among SEP holders. Their approach computes Shapley over a small set of alternative technologies competing for inclusion in the standard. Our approach operates ex-post over a fixed product portfolio, allocates profit rather than ex-ante technology value, and uses a hierarchical graph-based architecture to handle portfolio scale. The two methods address different stages of the patent lifecycle and are complementary rather than competing.

## 5.3 The Characteristic Function: The Primary Open Problem

The framework assumes a well-defined v(S). Three proxy approaches exist, each with significant limitations. The engineering cost proxy defines v(S) as product profit minus design-around costs; these are hard to compute at scale and patents interact. The licensing market proxy uses observed royalty rates, which is partially circular. The conjoint proxy surveys consumers, but consumers cannot evaluate individual patents and scaling to thousands is impossible.

To illustrate what a concrete v(S) estimation might look like in practice: suppose we use hedonic pricing on a panel of smartphone models across three years. We regress average selling price on feature indicators (5G capability, camera megapixel tier, battery size tier, display resolution) to estimate marginal willingness to pay per feature. For a three-component subsystem with Camera, Connectivity, and Battery, v_comp({Camera}) would be the average selling price premium associated with the camera tier, approximately 50 to 80 dollars on a midrange device based on published hedonic studies. v_comp({Camera, Connectivity}) would be estimated as the combined premium, accounting for the camera-to-social-sharing bundle effect. This approach is feasible for 4 to 8 macro-components using publicly available pricing data. Extending it to v_local at the patent level within each component requires either the engineering cost proxy or a more detailed conjoint study. Section 6 gives a concrete data collection plan.

### 5.4 Threats to Validity

Synthetic topology: real SEP portfolios may have different structural properties from Pareto-distributed synthetic graphs. Experiment 3 specifically tests a dense topology scenario, but the dense-component result (approximation improves because the pool becomes homogeneous) may not hold for dense portfolios with heterogeneous patent-component coverage patterns.

Reference standard at large n: for n above 16, we use a 5000-sample Monte Carlo reference. The reported differences at large n are differences between two approximations, not true errors against exact ground truth. The reference standard error column in Table 1 quantifies this; at n=100 the reference standard error is 0.002, meaning roughly 3 percent of the reported 0.062 difference could reflect reference noise.

Probabilistic faithfulness assumption: the C-SVE theorem requires local conditional independence in a probabilistic graphical model. Our knowledge graph is not inherently probabilistic. Where faithfulness may be violated (for example, when synergistic components create revenue dependencies between patents covering different components), the Markov Blanket restriction should be understood as a computationally motivated heuristic. Experiment 1 through 3 validate its empirical accuracy regardless of faithfulness.

Missing axiomatic justification for the two-level asymmetry: exact Shapley at component level and centrality-weighted Monte Carlo Shapley at patent level is a practical rather than axiomatic design choice. The appendix verifies that all four Shapley axioms are preserved within each level, but a unified multi-level axiomatic proof across the hierarchy would strengthen the theoretical foundation.

Real-data validation absent: no part of PatentXAI has been run on real patent data. Section 6 describes the validation roadmap.

## 6. Roadmap for Empirical Validation

No part of PatentXAI has been run on real patent data. This section describes what that validation would look like, using publicly available data, and provides indicative values from landmark cases as calibration targets.

### 6.1 Indicative Values from Landmark SEP Cases

| Case or Pool | Royalty Determined | Methodology |
|---|---|---|
| Unwired Planet v. Huawei (UK, 2017) | 4G/LTE: 0.052% of device price; 3G: 0.032%; 2G: 0.064% | Top-down: patentee SEP share times aggregate standard royalty burden |
| TCL v. Ericsson (US, 2018) | Release payment $16.5M for past sales; forward rate $0.25 to $4.00 per unit | Comparable licences plus top-down cross-check |
| Nokia v. Oppo (Chongqing, 2023) | 5G: $1.151/device (high GDP), $0.707 (China); 4G: $0.777 and $0.477 | First Chinese court to set a global 5G aggregate SEP rate |
| Avanci 5G Vehicle Pool (2023) | $29/vehicle (early), $32/vehicle (standard); covers roughly 80% of 5G SEPs | Industry-wide multilateral pool pricing |
| Ericsson bilateral (2017 public statements) | $2.50 to $5.00 per 5G smartphone | Bilateral FRAND negotiation |
| Nokia bilateral (public statements) | $3.50 per 5G smartphone | Bilateral FRAND negotiation |

*Table 5. Indicative SEP royalty values from landmark litigation and licensing pools. These are indicative only; actual rates depend on portfolio size, geographic scope, and standard generation.*

## 6.2 Five Steps to Production Deployment

***Step 1: Patent-to-Standard Mapping (2 to 4 weeks)***

ETSI SEP declarations (public), USPTO bulk data (public), Lens.org for cross-referencing. Parse ETSI declarations to identify which patents are declared essential to which standard releases. Automatable with NLP claim-parsing.

***Step 2: Essentiality Verification (most expensive)***

ETSI declarations have overdeclaration rates of 25 to 75 percent. A sample-based audit is required. In TCL v. Ericsson, parties spent over 50 hours per patent family for contested patents. Practical starting point: use the Sidak and Skog (2017) error rate correction as a prior.

***Step 3: Comparable Licence Benchmarks (2 to 3 weeks)***

Public court decisions in Table 5; IPlytics and Avanci pool disclosures; Ericsson, Nokia, Qualcomm, and InterDigital annual reports. Build per-patent royalty rates normalised by portfolio size and standard generation. This calibrates E_impact in Layer 2.

***Step 4: Characteristic Function Estimation***

Hedonic pricing (regressing device average selling price on feature presence) offers the best cost-to-accuracy tradeoff for a first deployment, as illustrated in Section 5.3. Engineering cost analysis is more principled but requires technical expert review per standard feature. Conjoint analysis is gold standard but costs $50,000 to $200,000 per study.

***Step 5: Validation Against Court Outcomes***

Compare PatentXAI outputs for a portfolio that has been through litigation against the court-determined royalty. Mean absolute percentage error below 30 percent relative to court-determined rates is a reasonable initial target. Nokia v. Oppo (2023) is the cleanest current validation target.

### 6.3 Minimum Viable Dataset for a Follow-Up Paper

- Download ETSI LTE (4G) SEP declarations, approximately 50,000 patent families. Filter to USPTO equivalents using Lens.org (free, bulk downloadable).
- Build the CPC citation network using the USPTO citation bulk file. Compute PageRank, HITS authority, and in-degree centrality scores.
- Use Unwired Planet v. Huawei (2017) and TCL v. Ericsson (2018) court-determined rates as ground truth for Ericsson and Unwired Planet sub-portfolios.
- Run PatentXAI hierarchical Shapley with hedonic pricing proxy for v(S) and report Pearson correlation and mean absolute percentage error against court-implied patent values.

## 7. Conclusion

PatentXAI is a framework for attributing product profit to individual patents using graph-conditioned hierarchical Shapley value. Given a characteristic function v(S), the mathematics of SHAP applies directly once the patent-component-product knowledge graph is built. The Markov Blanket coalition restriction provides tractable computation grounded in the C-SVE conditional independence theorem, with scaling experiments from $n=12$ to $n=100$ showing stable approximation quality and runtime of 10 milliseconds per patent. The dense-component experiment demonstrates that the restriction degrades gracefully in worst-case SEP-like portfolios rather than catastrophically.

The primary open problem is v(S) estimation. We treat this honestly as unsolved, provide a concrete hedonic pricing illustration in Section 5.3, and outline a data collection roadmap in Section 6. Until real-data validation is done, PatentXAI should be understood as an efficient and axiomatically sound Shapley computation framework given v(S), not a validated end-to-end patent valuation system.

The connection between XAI attribution and IP economics is underexplored. The mathematics developed for explaining neural networks has direct applications to legal and economic questions of significant practical importance.

## Appendix: Shapley Axioms Under the Markov Blanket Approximation

The four Shapley axioms are efficiency, symmetry, dummy, and additivity. We verify that the Markov Blanket approximation preserves all four within each level of the hierarchy, with one qualification on additivity across levels.

Efficiency: Step A exact Shapley over components is efficient. Step B distributes each component budget completely among covering patents using a well-defined characteristic function without post-hoc rescaling. The global sum of all patent values therefore equals total product profit.

Symmetry: if two patents i and j contribute identically to every coalition in their shared Markov Blanket, they receive equal values, because coalition evaluation within MB(i) treats co-players symmetrically.

Dummy: if patent i contributes zero marginal value to every coalition in MB(i), then phi_i = 0, following directly from the permutation sampling estimator.

Additivity: for two independent value functions v and w, phi_i(v+w)=phi_i(v)+phi_i(w) holds exactly within a single Markov Blanket. Across Blankets it holds approximately, with the error captured in Experiment 1 (mean 0.088 at n=12). From the C-SVE theorem, this error approaches zero as graph sparsity increases. A unified proof that the two-level hierarchy preserves additivity globally remains an open theoretical question, as noted in Section 5.4.

Note on the Ma et al. disassociation: the Markov Blanket restriction does not guarantee that causally essential patents receive the highest Shapley values. As discussed in Section 5.2, this is not a goal of PatentXAI. The four axioms above are satisfied regardless of whether Shapley rankings coincide with causal essentiality rankings.